\documentstyle[12pt,epsfig]{article}

\newcommand{\be}{\begin{equation}}
\newcommand{\ee}{\end{equation}}

\begin{document}
\topmargin 0pt
\oddsidemargin=0.5truecm
\renewcommand{\thefootnote}{\fnsymbol{footnote}}
\newpage
\setcounter{page}{0}
\begin{titlepage}
\vspace*{-2.0cm}
\begin{flushright}
hep-ph/0510272
\end{flushright}
\vspace*{0.1cm}
\begin{center}
{\Large \bf Neutrino Mass Matrices with Vanishing Determinant and $\theta_{13}$
} \\
\vspace{0.6cm}

\vspace{0.4cm}
{\large
Bhag C. Chauhan\footnote{On leave from Govt. Degree College, Karsog (H P)
India 171304. \\
E-mail: chauhan@cftp.ist.utl.pt},
Jo\~{a}o Pulido\footnote{E-mail: pulido@cftp.ist.utl.pt}}\\
\vspace{0.15cm}
{  {\small \sl Centro de F\'{\i}sica Te\'{o}rica das Part\'{\i}culas (CFTP) \\
 Departamento de F\'{\i}sica, Instituto Superior T\'{e}cnico \\
Av. Rovisco Pais, P-1049-001 Lisboa, Portugal}\\
}
\vspace{0.25cm}
and \\
\vspace{0.25cm}
{\large Marco Picariello\footnote{E-mail: Marco.Picariello@le.infn.it}} \\
\vspace{0.15cm}
{\small \sl  I.N.F.N. - Milano, and Dip. di Fisica, Universit\`a di Milano\\
and \\
Dipartimento di Fisica, Universit\`a di Lecce\\
Via Arnesano, ex Collegio Fiorini, I-73100 Lecce, Italia}
\end{center}
\vglue 0.6truecm
\begin{abstract}

We investigate the prospects for scenarios with vanishing determinant neutrino 
mass matrices and vanishing $\theta_{13}$ mixing angle. Normal and inverse
mass hierarchies are considered separately. For normal hierarchy it is found 
that neutrinoless double beta decay cannot be observed by any of the present
or next generation experiments. For inverse hierarchy the neutrinoless double
beta decay is, on the contrary, accessible to experiments.
We also analyse for both hierarchies the case for texture zeros and equalities
between mass matrix elements. No texture zeros are found to be possible nor 
any such equalities, apart from the obvious ones.
\end{abstract}

\end{titlepage}
\renewcommand{\thefootnote}{\arabic{footnote}}
\setcounter{footnote}{0}

\section{Introduction}
It is a well known fact that the neutrino mass matrix contains nine 
parameters while feasible experiments can hope to determine only seven 
of them in the foreseeable future. This situation can however be overcome,
with the number of free parameters being reduced, if physically 
motivated assumptions are made to restrict the form of the matrix. Among 
the most common such assumptions and as an incomplete list one may refer the
texture zeros \cite{Frampton:2002yf}, hybrid textures \cite{Kaneko:2005yz}, 
traceless condition \cite{He:2003rm}, \cite{He:2003nt}, \cite{Rodejohann:2003ir} 
and vanishing determinant \cite{Branco:2002ie}.
The former assumptions 
can be basis independent under certain conditions as shall be seen for the
traceless condition.
The latter being equivalent to one vanishing neutrino mass.

In this paper we perform the investigation of vanishing determinant neutrino
masses with vanishing $\theta_{13}$ \footnote{The $2\sigma$ range recently
obtained for this quantity is \cite{Fogli:2005cq} $sin^2 \theta_{13}=
0.9\pm^{2.3}_{0.9}\times 10^{-2} eV^2$, the lower uncertainty being purely formal,
corresponding to the positivity constraint $sin^{2}_{13}\geq 0$.}. We will 
assume that neutrinos are 
Majorana \cite{Kayser:2005cy}, as favoured by some experimental evidence
\cite{Klapdor-Kleingrothaus:2005zk}, and study the neutrino mass matrix $M$ 
in the weak basis where all charge leptons are already diagonalized. This 
is related to the diagonal mass matrix $D$ through the unitary transformation
\be
D=U^{T}_{MNS}MU_{MNS}
\ee
where we use the standard parametrization \cite{PDG}
\be
U_{MNS}=\!\!\left(\begin{array}{ccc}
c_{12}c_{13} & s_{12}c_{13} & s_{13}e^{-i\delta}\\
-s_{12}c_{23}-c_{12}s_{23}s_{13}e^{i\delta} & c_{12}c_{23}-s_{12}s_{23}s_{13}e^{i\delta}&
s_{23}c_{13} \\
s_{12}s_{23}-c_{12}c_{23}s_{13}e^{i\delta} & -c_{12}s_{23}-s_{12}c_{23}s_{13}e^{i\delta}&
c_{23}c_{13}\\
\end{array}\right).
\ee
where $\delta$ is a Dirac CP violating phase. Equation (1) is equivalent to
\be
M=U^{*}diag(m_{1},m_{2}e^{i\phi_1},m_{3}e^{i\phi_2})U^{\dagger}
\ee
where $\phi_{1},~\phi_{2}$ are two extra CP violating Majorana phases and 
$D=diag(m_{1},m_{2}e^{i\phi_1},m_{3}e^{i\phi_2})$. Applying
determinants properties
\be
\begin{array}{ll}
det~M&=det~(U^{*}DU^{\dagger}) \\
&=det~(U^{*}U^{\dagger}D) \\
&=det~U^{*}\ det~U^{\dagger}\ det~D \\
&=det~D~~(U~real) \\
&\ne det~D~~(U~complex)
\end{array}
\ee
because if matrix $U$ is real, $U^{*}U^{\dagger}=UU^{T}=1$, which is satisfied provided 
$\delta=0$ or $\theta_{13}=0$ (see eq.(2)).
Thus the determinant is not in general basis independent. In order
that $det~D=det~M$ it is necessary and sufficient that there is either no Dirac CP violation
or that it is unobservable. 
The same arguments hold for the condition $TrD=TrM$ \cite{He:2003nt}.

From eq. (4) we get that $det~M=0$ if and only if $det~D=0$,
because $det U^{\dagger}$ and $det U^{*}$ are not zero.
The vanishing determinant condition is basis independent, 
corresponding to a zero eigenvalue of the mass matrix.
So requiring $det~M=0$ is equivalent to assuming one of the neutrinos
to be massless.
This is realized for instance in 
the Affleck-Dine scenario for leptogenesis \cite{Affleck:1984fy},\cite{Murayama:1993em},
\cite{Dine:1995kz} which requires the lightest neutrino to be practically massless
($m\simeq 10^{-10} eV$) \cite{Asaka:2000nb},\cite{Fujii:2001sn}. Furthermore,
since in this paper we consider $\theta_{13}=0$,
the Dirac phase is unobservable and
the usual definition $U_{MNS}=U_{23}U_{13}U_{12}$ 
\cite{King:2002gx} simplifies to $U_{MNS}=U_{23}U_{12}$ with
\be
U_{23}=\left(\begin{array}{ccc} 1  &  0  &  0 \\
0   &  \alpha_{22}  &    \alpha_{23}   \\
0   &  \alpha_{32}  &    \alpha_{33}    \end{array}\right),
~U_{12}=\left(\begin{array}{ccc} \beta_{11}  &  \beta_{12}  &  0 \\
\beta_{21}   &  \beta_{22}  &    0   \\
0   &    0  &    1  \\  \end{array}\right) 
\ee
where the unitarity condition $(|\alpha_{22}\alpha_{33}-\alpha_{32}\alpha_{23}|=
|\beta_{11}\beta_{22}-\beta_{12}\beta_{21}|=1)$
implies $\alpha_{22}\alpha_{33}\alpha_{32}\alpha_{23}<0$ and 
$\beta_{11}\beta_{22}\beta_{12}\beta_{21}<0$ with
$\alpha_{22}=\pm cos \theta_{\otimes}$, $\beta_{11}=\pm cos \theta_{\odot}$,
the remaining matrix elements being evident. For neutrino masses and
mixings we refer to the following $2\sigma$
ranges~\cite{Fogli:2005cq,Gonzalez-Garcia:2004jd}
\be
\Delta m^2_{\odot}=m^2_{2}-m^2_{1}=7.92 \times 10^{-5}(1\pm 0.09) eV^2,
\ee
\be
\Delta m^2_{\otimes}=m^2_{3}-m^2_{2}=\pm 2.4 \times 10^{-3}(1\pm^{0.21}_{0.61}) eV^2
\ee
\be
sin^2\theta_{\odot}=0.314(1\pm^{0.18}_{0.15}),
\ee
\be
sin^2\theta_{\otimes}=0.44(1\pm^{0.41}_{0.22})
\ee
obtained from a 3 flavour analysis of all solar and atmospheric data. This favours
the widely used form of the $U_{MNS}$ matrix \cite{Scott:1999bs} (all entries taken 
in their moduli)
\be
U_{MNS}=\left(\begin{array}{ccc} \sqrt{\frac{2}{3}}&   \frac{1}{\sqrt{3}} & 0 \\
\frac{1}{\sqrt{6}} &   \frac{1}{\sqrt{3}} & \frac{1}{\sqrt{2}} \\
\frac{1}{\sqrt{6}} &   \frac{1}{\sqrt{3}} & \frac{1}{\sqrt{2}} \\
\end{array}\right).
\ee

The paper is organized as follows: in section 2 we derive all possible forms
of the mass matrix $M$ in this scenario for both normal and inverse 
hierarchies and investigate their consequences for $0\nu\nu\beta\beta$ decay. 
Since one of the neutrinos is massless, there is only one Majorana phase to 
be considered. In section 3 we investigate the prospects for texture zeros 
and equalities among matrix elements in both hierarchies and in section 4 we 
briefly expound our main conclusions.

\section{Mass matrices with $detM=0$ and $\theta_{13}=0$}

\subsection{Normal hierarchy (NH)}  

This is the case where the two mass eigenstates involved in the solar oscillations
are assumed to be the lightest so that $\Delta m^2_{\otimes}=\Delta m^2_{32}>0$.
We will consider this case as a departure from the degenerate one with 
$\Delta m^2_{\odot}=\Delta m^2_{21}=0$ and break the degeneracy with a real 
parameter $\epsilon$. Matrix $D$ with $m$ and $\epsilon$ both real is therefore
\be
D=diag(0,3 \epsilon e^{i\phi},m)
\ee
where $\phi$ is the Majorana relative phase between the second and third diagonal 
matrix elements ($\phi=\phi_1-\phi_2$ in the notation of section 2) and 
$\Delta m^2_{\odot}=9 \epsilon^2$. Using eqs.(5) the matrix $M$ is
\be
M\!\!=\!\!U_{23}U_{12}DU_{12}^T U_{23}^T\!\!=\!\!\left(\begin{array}{ccc}
3\epsilon e^{i\phi} \beta^2_{12} & 3\epsilon e^{i\phi} \alpha_{22}\beta_{12}\beta_{22} &
3\epsilon e^{i\phi} \alpha_{32}\beta_{12}\beta_{22} \\
3\epsilon e^{i\phi} \alpha_{22}\beta_{12}\beta_{22} &
3\epsilon e^{i\phi} \alpha_{22}^2 \beta_{22}^2 \!\!+\!\!m\alpha_{23}^2 &
3\epsilon e^{i\phi} \alpha_{22} \alpha_{32} \beta_{22}^2 \!\!+\!\!m\alpha_{23} \alpha_{33} \\
3\epsilon e^{i\phi} \alpha_{32}\beta_{12}\beta_{22} &
3\epsilon e^{i\phi} \alpha_{22} \alpha_{32} \beta_{22}^2 \!\!+\!\!m\alpha_{23} \alpha_{33} &
3\epsilon e^{i\phi} \alpha_{32}^2 \beta_{22}^2 \!\!+\!\!m\alpha_{33}^2 \\ 
\end{array}\right).
\ee
Owing to the sign ambiguities of parameters $\alpha$ and $\beta$, four possibilities
for matrix $M$ arise. 
Suppose entries 12 and 13 in this matrix have (+) (+) signs. Then $\alpha_{22},
\alpha_{32}$ have the same sign as $\beta_{12} \beta_{22}$, that is $\alpha_{22}
\alpha_{32}$ in the (23) entry is (+), implying the opposite sign for the
coefficient of $m$ ($\alpha_{23}\alpha_{33}$). 
So eq.(12) has the form 
\be
M=\left(\begin{array}{ccc} \epsilon e^{i\phi} &   \epsilon e^{i\phi} & \epsilon  e^{i\phi}\\
\epsilon e^{i\phi} & (m/2)+\epsilon e^{i\phi} &  -(m/2)+\epsilon  e^{i\phi}\\
\epsilon e^{i\phi} & -(m/2)+\epsilon e^{i\phi} &  (m/2)+\epsilon  e^{i\phi}\\ \end{array}\right)
\ee
Suppose entries 12 and 13 in the matrix have (-) (-) signs. Then $\alpha_{22},
\alpha_{32}$ have opposite sign to $\beta_{12} \beta_{22}$, that is they have
the same sign, so $\alpha_{22}\alpha_{32}$ is (+) and $\alpha_{23}\alpha_{33}$
is (-) so
\be
M=\left(\begin{array}{ccc} \epsilon  e^{i\phi}&   -\epsilon e^{i\phi} & -\epsilon  e^{i\phi}\\
-\epsilon e^{i\phi} & (m/2)+\epsilon  e^{i\phi}&  -(m/2)+\epsilon e^{i\phi} \\
-\epsilon e^{i\phi} & -(m/2)+\epsilon e^{i\phi} &  (m/2)+\epsilon e^{i\phi} \\ \end{array}\right)
\ee
Suppose entries 12 and 13 in the matrix have (+) (-) signs. Then $\alpha_{22},
\alpha_{32}$ have opposite signs to each other, so $\alpha_{22}\alpha_{32}$ is (-)
and $\alpha_{23}\alpha_{33}$ is (+). Hence
\be
M=\left(\begin{array}{ccc} \epsilon e^{i\phi} &   \epsilon e^{i\phi} & -\epsilon e^{i\phi} \\
\epsilon e^{i\phi} & (m/2)+\epsilon e^{i\phi} &  (m/2)-\epsilon e^{i\phi} \\
-\epsilon e^{i\phi} & (m/2)-\epsilon e^{i\phi} &  (m/2)+\epsilon  e^{i\phi}\\ \end{array}\right)
\ee
Suppose entries 12 and 13 in the matrix have (-) (+) signs. Then $\alpha_{22},
\alpha_{32}$ have opposite signs to each other, so $\alpha_{22}\alpha_{32}$ is (-)
and $\alpha_{23}\alpha_{33}$ is (+). Hence the matrix is
\be
M=\left(\begin{array}{ccc} \epsilon e^{i\phi} &   -\epsilon e^{i\phi} & \epsilon  e^{i\phi}\\
-\epsilon  e^{i\phi}& (m/2)+\epsilon  e^{i\phi}&  (m/2)-\epsilon  e^{i\phi}\\
\epsilon e^{i\phi} & (m/2)-\epsilon  e^{i\phi}&  (m/2)+\epsilon  e^{i\phi}\\ \end{array}\right)
\ee
All matrices (13), (14), (15), (16) have vanishing determinant as can be easily verified. 
For $0\nu\nu\beta\beta$ decay
\be
<m_{ee}>=|U_{e1}^2m_{1}+U_{e2}^2m_{2}e^{i\phi_1}+U_{e3}^2m_{3}e^{i\phi_2}| 
\ee
hence, for vanishing $m_1$ and $U_{e3}=s_{13}e^{-i\delta}$
\be
<m_{ee}>=|U_{e2}^2m_{2}e^{i\phi_1}|=\frac{1}{3}3\epsilon=
\frac{1}{3}\sqrt{\Delta m^2_{\odot}}\simeq 3\times 10^{-3}eV
\ee
where we used $\epsilon=\frac{1}{3}\sqrt{\Delta m^2_{\odot}}$. So the
Majorana phase is not an observable. 

There is no commonly accepted evidence in favour of $0\nu\nu\beta\beta$ decay  
but there exist reliable upper limits on $<m_{ee}>$ 
\be
<m_{ee}>\leq (0.3-1.2)eV~\cite{Klapdor-Kleingrothaus:2005zk},~~
<m_{ee}>\leq (0.2-1.1)eV~\cite{Arnaboldi:2005cg}
\ee
where the uncertainties follow from the uncertainties in the nuclear matrix
elements. The future CUORE experiment \cite{Capelli:2005jf}, of which 
CUORICINO is a test version \cite{Arnaboldi:2005cg}, is expected to improve
this upper bound to $3\times 10^{-2}eV$. Other experiments are also proposed 
(MAJORANA \cite{Aalseth:2005mn}, GENIUS \cite{Klapdor-Kleingrothaus:1999hk}, 
GEM \cite{Zdesenko:2001ee} and others) in which the sensitivity of a few 
$10^{-2}eV$ is planned to be reached.

Conclusion: {\it vanishing determinant with vanishing $\theta_{13}$ and
NH implies that $0\nu\nu\beta\beta$ decay cannot be detected 
even in the next generation of experiments. This remains unchanged even
if $\theta_{13}\ne 0$, since the largest mass ($m_3$) multiplies $s_{13}^2$ 
in eq.(17).} 

\subsection{Inverse Hierarchy (IH)} 

We start with matrix $D$ in the form $D=diag\{m,(m+\epsilon)e^{i\phi},0\}$
where $m$, $\epsilon$ are complex, $|m|\simeq \sqrt{\Delta m^{2}_{\otimes}}$,
$|\epsilon| \simeq \sqrt{\Delta m^{2}_{\odot}}$ and chosen in such
a way that $m+\epsilon=\tilde m$ is real  
($\epsilon=0$ corresponds to the degenerate case). Alternatively
$D=diag\{\tilde m- \epsilon, \tilde m e^{i\phi},0\}$
with, of course,
$\tilde m- \epsilon$ complex. Multiplying the whole matrix by the inverse 
phase of $\tilde m- \epsilon$, it can be redefined as 
\be
D=diag\{\tilde m- \lambda, \tilde m e^{i(\phi-\psi)},0\}
\ee
with $\lambda$ real
and defined by $(\tilde m- \epsilon)e^{-i\psi}= \tilde m- \lambda$ (notice that 
$\tilde m- \epsilon=|\tilde m- \epsilon|e^{i\psi}$ and $\tilde m =
\sqrt{\Delta m^{2}_{\otimes}}$). There are two solutions for $\lambda$. In fact,
imposing the solar mass square difference
\be 
\Delta m^{2}_{\odot}=|d_{22}|^2-|d_{11}|^2=\tilde m^2- \tilde m^2 +2\lambda \tilde {m}-
\lambda^2
\ee
and solving the quadratic equation $\lambda^2 - 2\lambda \tilde m +
\Delta m^{2}_{\odot}=0$ one gets
\be
\lambda=\tilde m\pm\sqrt{\tilde m^2- \Delta m^{2}_{\odot}}=\lambda_{\pm}.
\ee
Notice that $\lambda_{+}$ is large and $\lambda_{-}$ is small. To first order 
in $\frac{\Delta m^2_{\odot}}{\tilde m^2}=\frac{\Delta m^2_{\odot}}
{\Delta m^2_{\otimes}}\simeq 0.30$ one has
\be
\lambda_{+}=\tilde m (2 -\frac{1}{2}\frac{\Delta m^2_{\odot}}{\Delta m^2_{\otimes}})
\simeq 1.85 \tilde m
\ee
\be
\lambda_{-}=\frac{\tilde m}{2}\frac{\Delta m^2_{\odot}}{\Delta m^2_{\otimes}}
\simeq \frac{\tilde m}{60}
\ee
It is straightforward to see that $D(\lambda_{-},\phi+ \pi)=-D(\lambda_{+},\phi)$
and the same property holds for matrix $M$, namely $M(\lambda_{-},\phi+ \pi)=
-M(\lambda_{+},\phi)$ because $U_{MNS}$ is invariant under the transformations
$\lambda_{+} \rightarrow \lambda_{-}$ and $\phi \rightarrow \phi +\pi$. 
So the two solutions for $\lambda$ are equivalent: one may take either
\be 
\lambda_{+}~,~\psi=0
\ee
or
\be
\lambda_{-}~,~\psi=\pi.
\ee
Using $M\!=\!U_{23}U_{12}DU_{12}^T U_{23}^T$ with eqs.(5), (20), the matrix $M$ 
has now the form
\be
M=\left(\begin{array}{ccc}
\tilde m(1-\frac{t}{3})-\frac{2}{3}\lambda & (sign)\frac{1}{3}(\tilde {m}t-\lambda) & 
(sign)\frac{1}{3}(\tilde {m}t-\lambda) \\
(sign)\frac{1}{3}(\tilde {m}t-\lambda) & \tilde m(\frac{1}{2}-\frac{t}{3})-\frac{\lambda}{6} &
(sign)[\tilde m(\frac{1}{2}-\frac{t}{3})-\frac{\lambda}{6}] \\
(sign)\frac{1}{3}(\tilde {m}t-\lambda) & (sign)[\tilde m(\frac{1}{2}-\frac{t}{3})-\frac{\lambda}{6}] &
\tilde m(\frac{1}{2}-\frac{t}{3})-\frac{\lambda}{6} \\
\end{array}\right). 
\ee
which also verifies $det~M=0$ as expected.
Equation (27) is formally the same for $\lambda=\lambda_{+}$ and $\lambda=\lambda_{-}$
with the definition $t=1-e^{i\phi}$ for $\lambda=\lambda_{+}$ and $t=1+e^{i\phi}$
for $\lambda=\lambda_{-}$, the sign affecting the exponential 
being related to the $\psi$ phase.
The structure of (+) and (-) in eq.(27) is the same as before ((13), (14), (15), (16)):
equal signs in entries $M_{12}$, $M_{13}$ correspond to (+) in both entries $M_{23}$, 
$M_{32}$ while different signs in $M_{12}$, $M_{13}$ correspond to (-) in both entries 
$M_{23}$, $M_{32}$.
Eq. (27) is the equivalent for IH of (13), (14), (15), (16) for NH. 

For $\beta \beta_{0\nu\nu}$ decay we have
\be
<m_{ee}>=|U_{e1}^2m_{1}+U_{e2}^2m_{2}e^{i\phi_1}+U_{e3}^2m_{3}e^{i\phi_2}|=
|\frac{2}{3}(\tilde m- \lambda_{\pm})\pm \frac{1}{3}\tilde m e^{i(\phi)}|. 
\ee
The quantity $m_{ee}$ is displayed in fig.1 as a function of the phase difference $\phi$.
The shaded areas correspond to the $1\sigma$ uncertainties in the solar angle $\theta_{\odot}$.
It is seen from eq.(28) and fig.1 that for inverse hierarchy (vanishing $\theta_{13}$ 
and mass matrix determinant) $\beta \beta_{0\nu\nu}$ decay is phase dependent and within 
observational limits of forthcoming experiments. So:

Conclusion: {\it models with vanishing determinant mass matrix and vanishing $\theta_{13}$
provide, in inverse hierarchy, a Majorana phase dependent $\beta \beta_{0\nu\nu}$ decay 
which is physically observable for most values of the phase 
in the next generation of experiments.}

\begin{figure}[h]
\setlength{\unitlength}{1cm}
\begin{center}
\hspace*{-1.6cm}
\epsfig{file=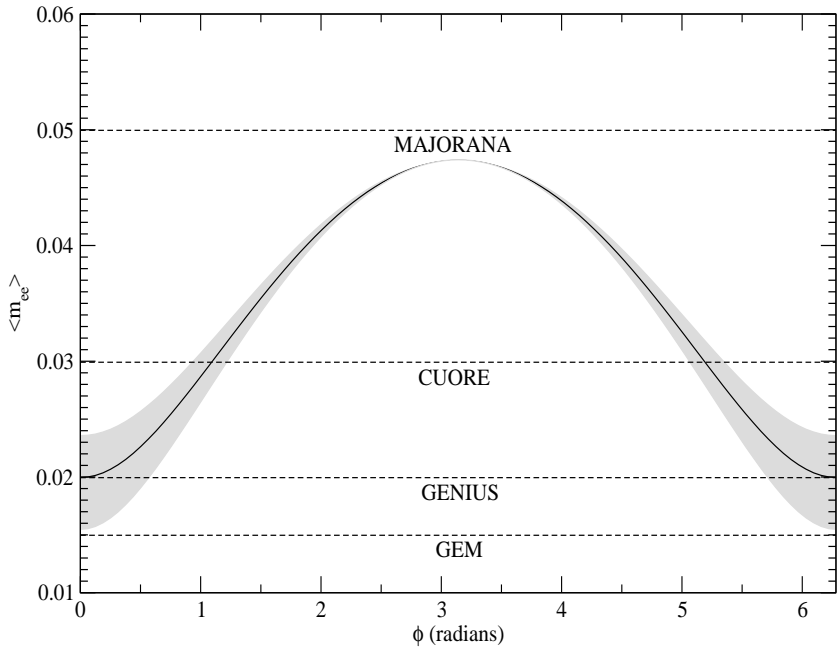,height=13.0cm,width=11.0cm,angle=270}
\end{center}
\caption{ \it$\beta \beta_{0\nu\nu}$ decay effective mass parameter $<m_{ee}>$ as a function 
of the Majorana phase $\phi$ showing its accessibility for forthcoming experiments.}  
\label{fig1}
\end{figure}

\section{Texture zeros and equalities between M matrix elements}

\subsection{Texture zeros}

Here we analyze the possibility of vanishing entries in the mass matrix $M$. 
Taking first NH and recalling eqs.(13)-(16), it is seen that this implies
either $\tilde m /2=\pm \epsilon e^{i \phi}$ or $\epsilon=0$, both situations 
being impossible. For IH three cases need to be considered: \\
$(a)~~M_{11}=0$ \\
We have in this case $\tilde m(1-\frac{t}{3})-\frac{2}{3}\lambda=0$ implying
\be
\tilde m(3-t)=2\lambda.
\ee
Replacing $t\rightarrow 1-e^{i\phi}$ and $\lambda \rightarrow \lambda_{+}$ this
leads to
\be
e^{i\phi}=2\sqrt{1-\frac{\Delta m^2_{\odot}}{\Delta m^2_{\otimes}}}
\ee
\vspace{0.3cm}
which is experimentally excluded. \\   
$(b)~~M_{12}=0$ \\
This gives $\tilde m t-\lambda=0$, hence using the same replacement
\be
e^{i\phi}=-\sqrt{1-\frac{\Delta m^2_{\odot}}{\Delta m^2_{\otimes}}}
\ee
\vspace{0.3cm}
which is also impossible since $\Delta m^2_{\odot}=0$ is strictly excluded experimentally. \\
$(c)~~M_{22}=0$ \\ 
This gives $\tilde m(\frac{1}{2}-\frac{t}{3})-\frac{\lambda}{6}=0$, hence using the same 
replacement 
\be
e^{i\phi}=\frac{1}{2}\sqrt{1-\frac{\Delta m^2_{\odot}}{\Delta m^2_{\otimes}}}
\ee
which is also experimentally excluded. In the former cases (a), (b), (c) the same 
results are of course obtained with the replacement $t\rightarrow 1+e^{i\phi}$ and 
$\lambda \rightarrow \lambda_{-}$, as can be easily verified. 
So zero mass textures are not possible in the present scenario. 

The same conclusion can be obtained using the results from the literature. In fact
the analytical study of various structures of the neutrino mass matrix was
presented systematically by Frigerio and Smirnov \cite{Frigerio:2002fb} who also discussed 
the case of equalities of matrix elements. Here we use a result from \cite{Xing:2003ic} 
where specific relations among the mixing angles were derived for one texture zero and 
one vanishing eigenvalue. We refer to table I of \cite{Xing:2003ic} and first to NH. 
Using their 
definition of parameter $\chi=\left|\frac{m_2}{m_3}\right|$ we have in our model $\chi=
\sqrt {\frac{\Delta m^2_{\odot}}{\Delta m^2_{\otimes}}} =0.182$ and so for 
cases A, B, C, D, E, F respectively in their notation
\be
\chi=0,~\chi=0,~\chi=0,~\chi=1.50,~\chi=1.50,~\chi=1.50 
\footnote{For instance for case D, normal hierarchy and $\theta_{13}\ne 0$ 
we have, with the best fit values given in \cite{Fogli:2005cq}, 
$\chi=\frac{0.436}{0.385-0.0437c_{\delta}}~~({\delta}-Dirac~phase).$}
\ee
For inverse hierarchy, defining  
$\eta=\frac{m_1}{m_2}=\frac{|\tilde m -\lambda_{\pm}|}{\sqrt{\Lambda m^2_{\otimes}}}=
\sqrt{1-\frac{\Delta m^2_{\odot}}{\Delta m^2_{\otimes}}}=0.983$ we have for cases 
A, B, C, D, E, F respectively
\be
\eta=0.50,~\eta=1,~\eta=1,~\eta=2.0,~\eta=2.0,~\eta=2.0
\ee
Notice that $0.953<\eta < 0.988$ (using $1\sigma$ upper and lower values for the solar 
and atmospheric mass square differences). So one can draw the following 

\vspace{0.6cm}
{\it Conclusion: both NH and IH cannot work with det D= det M=0, vanishing $\theta_{13}$
and one texture zero. In other words, vanishing determinant scenarios with $\theta_{13}=0$
are experimentally excluded, unless they have no texture zeros}.

\subsection{Equalities between matrix elements}

First we consider the case of NH. Equations (13)-(16) can be written in the general form
\be
M=\left(\begin{array}{ccc} \epsilon e^{i\phi} &  sign(\epsilon e^{i\phi}) & sign(\epsilon  e^{i\phi})\\
sign(\epsilon e^{i\phi}) & (m/2)+\epsilon e^{i\phi} & sign[-(m/2)+\epsilon e^{i\phi}]\\
sign(\epsilon e^{i\phi}) & sign[-(m/2)+\epsilon e^{i\phi}] &  (m/2)+\epsilon  e^{i\phi}\\ \end{array}\right)
\ee
and using the same sign conventions as in eqs.(13)-(16), it is seen that $|M_{11}|=|M_{12}|$, 
$|M_{12}|=|M_{13}|$, $M_{22}=M_{33}$. Hence the
relations to be investigated are $M_{11}=M_{22}$, $|M_{11}|=|M_{23}|$, $|M_{22}|=|M_{23}|$.

Equation $M_{11}=M_{22}$ implies $m=0$ which is impossible.

Equation $|M_{11}|=|M_{23}|$ yields
\be (a)~~\epsilon e^{i\phi}=(-m/2)+\epsilon e^{i\phi} 
\ee
leading to $m=0$, and
\be (b)~~\epsilon e^{i\phi}=(+m/2)-\epsilon e^{i\phi} 
\ee
leading to $\frac{4}{3}=\sqrt{1+\frac{\Delta m^2_{\otimes}}{\Delta m^2_{\odot}}}$
which is also experimentally excluded.

Equation $|M_{22}|=|M_{23}|$ yields
\be (a)~~(+m/2)+\epsilon e^{i\phi}=(-m/2)+\epsilon e^{i\phi}
\ee
leading to $m=0$, and 
\be (a)~~(+m/2)+\epsilon e^{i\phi}=(+m/2)-\epsilon e^{i\phi}
\ee
leading to $\epsilon=0$, both experimentally excluded.

Next we consider IH. We use eq. (27) and note that the matrix is symmetric, so there are
at first sight 6 independent entries. However $M_{22}=M_{33}$, $|M_{12}|=|M_{13}|$,
$|M_{22}|=|M_{23}|$, so there remain 3 independent matrix elements and therefore 3 
equalities to be investigated.  
In other words, there are three different moduli only: $|M_{11}|,
~|M_{12}|,~|M_{22}|$. So the three equalities to be investigated are
$|M_{11}|=|M_{12}|$, $|M_{11}|=|M_{23}|$, $|M_{12}|=|M_{23}|$. 

Equality $|M_{11}|=|M_{12}|$ yields two cases
\be (a)~~\tilde m(1-\frac{t}{3})-\frac{2}{3}\lambda=\frac{1}{3}(\tilde m t-\lambda)
\ee
which upon using $\lambda=\lambda_{\pm}$ for $t=1\mp e^{i\phi}$ gives  
\be
\tilde m -\lambda_{\pm}={\mp}2 \tilde m e^{i\phi}
\ee
which is impossible to satisfy, as seen from eq.(22), and 
\be
(b)~~\tilde m(1-\frac{t}{3})-\frac{2}{3}\lambda=-\frac{1}{3}(\tilde m t-\lambda)
\ee
leading to 
\be
\tilde m=\lambda,
\ee
also impossible, eq.(22).

Equality $|M_{11}|=|M_{23}|$. The two cases to be considered are
\be
(a)~~\tilde m(1-\frac{t}{3})-\frac{2}{3}\lambda= \tilde m(\frac{1}{2}-\frac{t}{3})-\frac{\lambda}{6}
\ee
from which
\be
\tilde m=\lambda 
\ee
which cannot be satisfied (eq.(22)) and
\be
(b)~~\tilde m(1-\frac{t}{3})-\frac{2}{3}\lambda=-\tilde m(\frac{1}{2}-\frac{t}{3})+\frac{\lambda}{6}
\ee
which upon using $\lambda=\lambda_{\pm}$ for $t=1\mp e^{i\phi}$ gives
\be
\tilde m -\lambda_{\pm}={\mp}\frac{4}{5}\tilde m e^{i\phi}
\ee
or equivalently
\be
5\sqrt{1-\frac{\Delta m^2_{\odot}}{\Delta m^2_{\otimes}}}=4e^{i\phi}
\ee 
which is cannot be satisfied even if $\phi=0$. (Maximizing $\Delta m^2_{\odot}$
and minimizing $\Delta m^2_{\otimes}$ (1 $\sigma$) the above square root verifies 
$0.953<\sqrt{1-\frac{\Delta m^2_{\odot}}{\Delta m^2_{\otimes}}}<0.988$).

Equality $|M_{12}|=|M_{23}|$. The two cases are now
\be
(a)~~\frac{1}{3}(\tilde m t-\lambda)=\tilde m(\frac{1}{2}-\frac{t}{3})-\frac{\lambda}{6}
\ee
which gives $\tilde m-\lambda_{\pm}=\pm 4\tilde m e^{i\phi}$ or $\pm \sqrt{1-\frac{\Delta m^2_{\odot}}
{\Delta m^2_{\otimes}}}=\pm 4e^{i\phi}$, again impossible, and
\be
(b)~~\frac{1}{3}(\tilde m t-\lambda)=-\tilde m(\frac{1}{2}-\frac{t}{3})-\frac{\lambda}{6}
\ee
or $\tilde m=\lambda$, also impossible.
All these impossibilities mean {\it experimentally excluded}.

Moreover, if $M_{12}$ and $M_{13}$ have opposite signs, since $|M_{12}|=|M_{13}|$, they both 
vanish, implying two texture zeros which is excluded. The same is true for $M_{22}$ and 
$M_{23}$. Recall that one texture zero with vanishing determinant cannot work with 
$\theta_{13}=0$ (see section 3.1). Hence:

\vspace{0.6cm}
{\it Conclusion: equalities between mass matrix elements apart from the obvious ones are 
experimentally excluded.}

\section{Conclusions}

We have investigated the prospects for neutrino mass matrices with vanishing determinant and
$\theta_{13}$. The vanishing determinant condition alone is expressed by two real conditions,
so the original nine independent parameters in these matrices are reduced to seven. Hence the
undesirable situation of existing and planned experiments not being able to determine all 
these nine quantities is in this case overcome. Furthermore, as shown in the introduction, 
the vanishing of $\theta_{13}$ implies that the CP violating Dirac phase is unobservable 
and the mass matrix can be diagonalized by a real and orthogonal matrix.
In such case the mass matrix determinant is basis independent, $det~M=det~D$, while the
vanishing determinant condition is always basis independent. So $det~M=0$ is always 
equivalent to the lightest neutrino being massless. 

We considered both the normal and inverse mass hierarchies.
Summarizing our main conclusions for vanishing determinant mass matrices with vanishing 
$\theta_{13}$:

In the case of normal hierarchy there can be no observable  
$\beta\beta_{0\nu\nu}$ decay. For inverse hierarchy 
$\beta\beta_{0\nu\nu}$ decay depends on the Majorana phase and can be observed in the 
next generation of experiments for all or most of the possible phase range.
 
Texture zeros and equalities between mass matrix elements besides the obvious ones are 
incompatible with experimental evidence.

\end{document}